\documentclass[
  10pt,
  aps,
  prb,
  superscriptaddress,
  twocolumn,
  floatfix,
  showpacs
]{revtex4-1}

\usepackage{afterpage}

\usepackage{color}

\usepackage{mathtools}

\usepackage{physics}

\newcommand{\etal}{\textit{et~al.}}

\newcommand{\ie}{i.\,e.}

\newcommand{\nn}[1]{\langle #1 \rangle}
\newcommand{\cre}[1]{#1^{\dagger}}
\newcommand{\ann}[1]{#1^{\mathstrut}}
\newcommand{\cnt}[1]{\cre{#1}\ann{#1}}

\newcommand{\vc}[1]{\mathrm{#1}}
\DeclareMathOperator{\Id}{I}
\newcommand{\I}{{\mathrm{i}\mkern1mu}}
\newcommand{\E}{{\mathrm{e}\mkern1mu}}

\newcommand*{\tran}{^{\mkern-1.5mu\mathsf{T}}}

\newcommand{\Z}{\mathcal{Z}}

\newcommand{\cl}{{\mathrm{cl}}}
\newcommand{\Ncl}{{N_{\mathrm{cl}}}}
\newcommand{\epscl}{{\bar{\epsilon}}}
\newcommand{\Kcl}{{\vb{K}}}
\newcommand{\kcl}{{\tilde{\vb{k}}}}
\newcommand{\rcl}{{\tilde{\vb{r}}}}

\usepackage[
  pdftitle={Spreading of correlations in the Falicov-Kimball model},
  pdfauthor={Andreas J. Herrmann and Andrey E. Antipov and Philipp Werner}
]{hyperref}

\usepackage[
  capitalise
]{cleveref}
\crefformat{table}{#2Tab.~\!#1#3}
\crefmultiformat{table}{Tabs.~#2#1#3}{\ and~#2#1#3}{, #2#1#3}{\ and~#2#1#3}
\Crefformat{table}{#2Tab.~\!#1#3}
\Crefmultiformat{table}{Tabs.~#2#1#3}{\ and~#2#1#3}{, #2#1#3}{\ and~#2#1#3}
\crefformat{figure}{#2Fig.~\!#1#3}
\crefmultiformat{figure}{Figs.~#2#1#3}{\ and~#2#1#3}{, #2#1#3}{\ and~#2#1#3}
\Crefformat{figure}{#2Fig.~\!#1#3}
\Crefmultiformat{figure}{Figs.~#2#1#3}{\ and~#2#1#3}{, #2#1#3}{\ and~#2#1#3}

\begin{document}

\title{%
  Spreading of correlations in the Falicov-Kimball model
}
\author{Andreas J. Herrmann}
\email[]{andreas.herrmann@unifr.ch}
\affiliation{%
  Department of Physics,
  University of Fribourg,
  1700 Fribourg,
  Switzerland
}
\author{Andrey E. Antipov}
\affiliation{%
  Station Q, Microsoft Research, Santa Barbara, California 93106, USA
}
\author{Philipp Werner}
\affiliation{%
  Department of Physics,
  University of Fribourg,
  1700 Fribourg,
  Switzerland
}
\date{\today}

\begin{abstract}
  We study dynamical properties
  of the one- and two-dimensional Falicov-Kimball model
  using lattice Monte Carlo simulations.
  In particular, we calculate the spreading of charge correlations
  in the equilibrium model and after an interaction quench.
  The results show a reduction of the light-cone velocity
  with interaction strength at low temperature,
  while the phase velocity increases.
  At higher temperature,
  the initial spreading is determined by the Fermi velocity
  of the noninteracting system
  and the maximum range of the correlations decreases
  with increasing interaction strength.
  Charge order correlations in the disorder potential
  enhance the range of the correlations.
  We also use the numerically exact lattice Monte Carlo results
  to benchmark the accuracy
  of equilibrium and nonequilibrium
  dynamical cluster approximation calculations.
  It is shown that
  the bias introduced by the mapping to a periodized cluster is substantial,
  and that from a numerical point of view,
  it is more efficient to simulate the lattice model directly.
\end{abstract}

\maketitle

\section{Introduction}

Physical systems are characterized by their response to external perturbations.
Slow or weak perturbations probe the equilibrium state of the system
through response functions,
which can be classified in terms of the low-energy excitations.
Outside of this regime
the nonequilibrium dynamics mixes excitations at different energy scales
and can be complicated.
In correlated quantum systems
non-equilibrium studies reveal a plethora of new phenomena
\cite{dalessio_quantum_2016,aoki_nonequilibrium_2014}
and novel states of matter.
\cite{basko_metalinsulator_2006,else_floquet_2016,zhang_observation_2017,martin_topological_2016}

A general classification
of universal features of non-equilibrium transport in quantum systems
is currently lacking.
Nevertheless, theoretical predictions exist for the spreading of correlations.
Lieb and Robinson \cite{lieb_finite_1972}
showed that for interactions of finite range,
there is a maximum velocity associated with this spreading.
The resulting light-cone dynamics
manifests itself in the commutators of observables,
which are related to physical response functions,
while anticommutators can exhibit algebraic tails
that extend beyond the light-cone.
\cite{medvedyeva_spatiotemporal_2013,abeling_analysis_2017}
In the case of a noninteracting Fermion model
the spreading velocity is determined by the maximum Fermi velocity.
The effects of interactions and disorder
modify this velocity and (in a localized phase)
may limit the range of the correlations.

\afterpage{%
\begin{figure}
  \includegraphics[width=0.95\linewidth]{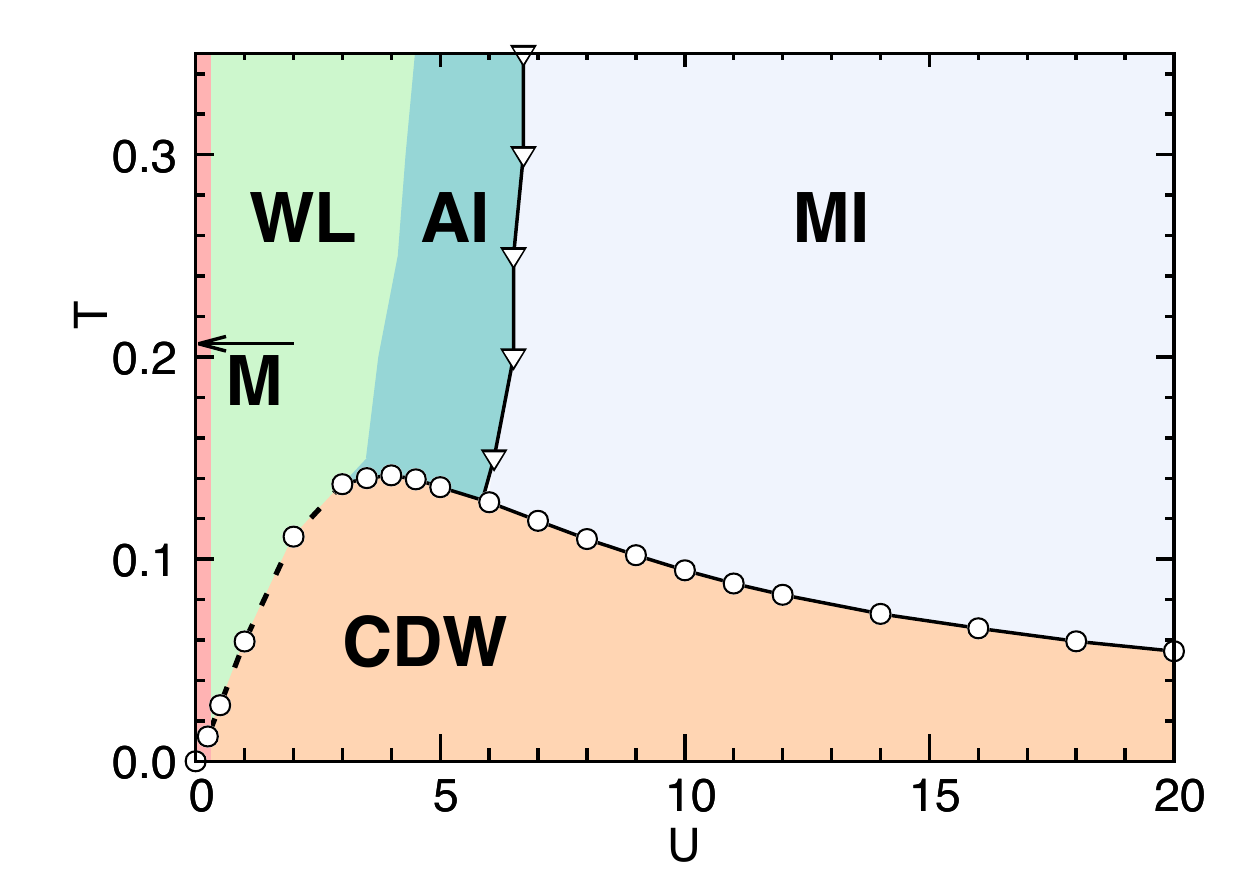}
  \caption{\label{fig:phase_diagram}%
    Phase-diagram of the spin-less particle-hole symmetric Falicov-Kimball model
    on a two-dimensional square lattice,
    in the space of local interaction $U$ and temperature $T$.
    The unit of energy is the nearest-neighbor hopping.
    The noninteracting system is metallic (M)
    while the interacting system at low temperatures
    is in a charge density wave (CDW) phase.
    At temperatures $\gtrsim 0.15$, the system goes through
    weakly localized (WL), Anderson insulator (AI), and Mott insulator (MI)
    phases with increasing $U$.
    The WL phase appears in finite size simulations (here $48\times 48$)
    and is replaced by the AI in the thermodynamic limit.
    (Lattice Monte Carlo results by Antipov~\etal
    \cite{antipov_interaction-tuned_2016})
  }
\end{figure}
}

A relatively simple model
which allows to explore the influence of disorder and correlations
on the light-cone dynamics is the Falicov-Kimball model.
\cite{freericks_exact_2003,falicov_simple_1969}
This model describes mobile $c$~electrons
which interact with immobile $f$~electrons.
In equilibrium, this interaction produces
a self-consistently determined disorder potential for the $c$~electrons.
Recent lattice Monte Carlo simulations
\cite{maska_thermodynamics_2006,antipov_interaction-tuned_2016}
showed that the half-filled Falicov-Kimball model
on the square lattice
exhibits a rich phase diagram with
metallic, weakly localized, Anderson insulating,
Mott-like insulating, and charge ordered insulating (CDW) phases,
see \cref{fig:phase_diagram}.
The metallic (M) and weakly localized (WL) phases are conducting.
Both the weakly localized and Anderson insulating (AI) phases
are characterized by a nonzero density of states at the Fermi level
and a finite localization length.
The crossover between the AI and WL regimes
occurs when the localization length reaches the system size.
In the thermodynamic limit,
the WL phase is replaced by the AI phase
with a vanishing static conductivity.
The Mott-like insulating (MI) and charge ordered insulating (CDW) phases
have no available states in the proximity of the Fermi energy,
either due to \textit{c-f} electronic interactions (MI)
or the presence of a static checkerboard order (CDW).

For a given $f$~configuration,
the dynamics of the noninteracting $c$~electrons can be computed explicitly.
Therefore the dynamics of the Falicov-Kimball model
can be studied using a sign-problem free lattice Monte Carlo simulation
of real-time correlation functions.

In this paper, we use this method to study time-dependent correlation functions
in the different equilibrium phases of the Falicov-Kimball model,
and after interaction quenches between different physical regimes.
We connect the maximum spatial extent of the spreading correlations
and their phase and group velocities
to characteristics of the underlying equilibrium phases.
Specifically, Anderson and Mott-like localized phases
show a finite range of correlations,
and an increasing (decreasing) phase (group)
velocity with increasing interaction strength.
The onset of charge order changes this behavior
and the correlations spread across the whole system.

We will furthermore use the numerically exact lattice Monte Carlo data
to benchmark the dynamical cluster approximation (DCA) results
in and out of equilibrium.
\cite{hettler_nonlocal_1998,hettler_dynamical_2000,herrmann_nonequilibrium_2016}
The dynamical mean field theory (DMFT)
and its cluster extensions
provide a numerically tractable tool for the study of nonequilibrium problems,
\cite{aoki_nonequilibrium_2014}
such as parameter quenches and electric field excitations.
\cite{tsuji_nonequilibrium_2014,eckstein_ultra-fast_2016,herrmann_nonequilibrium_2016}
However, relatively little is currently known
about the effect of nonlocal correlations on the nonequilibrium properties
of perturbed lattice systems, and on the ability of cluster DMFT methods
to capture this physics.
In the case of the one-dimensional Hubbard model,
a systematic comparison between cluster DMFT and time-dependent DMRG
has been presented in Ref.~\onlinecite{tsuji_nonequilibrium_2014}
for interaction quenches from a noninteracting initial state.
It was found that the dynamics of local observables converges much faster
with cluster size than the dynamics of nonlocal observables,
and that a proper averaging over different cluster geometries
can considerably improve the convergence with cluster size.
The use of a weak-coupling perturbative impurity solver
however limited these benchmarks to the weakly correlated regime.
In the case of the Falicov-Kimball model,
the cluster impurity problem can be solved exactly,
so that the nonequilibrium dynamics can be simulated
for arbitrary interaction strength.
\cite{herrmann_nonequilibrium_2016}
This, in combination with exact Monte Carlo benchmark results,
allows us to reveal the errors introduced by neglecting long-range correlations,
or by assuming translational symmetry on the cluster.

The paper is organized as follows.
Section II details the Monte Carlo simulation method
and the dynamical cluster approximation for the Falicov-Kimball model.
Section IIIA investigates the convergence
of the lattice Monte Carlo and DCA simulations with system size,
and provides benchmark results for equilibrium and nonequilibrium DCA.
Section IIIB investigates the spreading of correlations
in the one-dimensional and two-dimensional Falicov-Kimball model,
while Section IV contains a discussion and outlook.

\section{Methods}
\label{sec:methods}

\subsection{Lattice Monte Carlo}
\label{subsec:lattice_monte_carlo}

The lattice Hamiltonian of the Falicov-Kimball model
in $d=1,2$ dimensions is given by
\begin{equation}
  H
  =
  -t \sum_{\langle i,j \rangle}  \cre{c_i} \ann{c_j}
  +\sum_i U \cnt{c_i} \cnt{f_i}
  -\sum_i \mu \qty( \cnt{c_i} + \cnt{f_i} )
  ,
  \label{eq:fkm_hamiltonian}
\end{equation}
where
$t$ is the hopping between nearest neighbor sites,
$U$ is the interaction parameter,
and $\mu$ is the chemical potential
($\mu=U/2$ corresponds to half-filling).
The $c$~electrons are itinerant,
while the $f$~electrons are localized,
and the number of both species is conserved.

In the Monte Carlo treatment,
we restrict the system to a $d$-dimensional cluster of linear size $L$
with periodic boundary conditions.
The partition function of the thermal system
at inverse temperature $\beta$
is given by
\begin{equation}
  \mathcal{Z}
  =
  \Tr \E^{-\beta H}
  =
  \sum_\alpha
  \E^{\beta\mu \sum_i n^f_{\alpha\,i}}
  \Tr_c \E^{-\beta H_\alpha}
  ,
\end{equation}
where the trace over the $f$~electron configurations enumerated by $\alpha$
is taken explicitly and $n^f_{\alpha\,i} \in \{0,1\}$
denotes the $f$~electron occupation
in configuration $\alpha$ at lattice site $i$.
$H_\alpha = \ev{H + \sum_i \mu \cnt{f_i}}{\alpha}=\underline c^\dagger \underline{H}_\alpha \underline c$
is the $c$~electron Hamiltonian
with a site-dependent potential term defined by $\alpha$.
Here $\underline c^\dagger$ is the vector of creation operators
(of size $S=L^d$)
and $\underline H_\alpha$ an $S\times S$ matrix.

Since $H_\alpha$ is quadratic in the fermionic operators,
we can express the partition function as
\begin{equation}
  \mathcal{Z}
  =
  \sum_\alpha
  P_\alpha
  =
  \sum_\alpha
  \E^{\beta\mu \sum_i n^f_{\alpha\,i}}
  \det \qty[ \Id + \E^{-\beta \underline H_\alpha} ]
  ,
\end{equation}
and we may interpret $P_\alpha$
as an unnormalized probability amplitude
for the $f$~electron configuration $\alpha$.

The thermal expectation value of an observable
$O=\underline c^\dagger \underline O \,\underline c$
can be calculated as
\begin{align}
  \ev{O}
  &=
  \frac{1}{\mathcal{Z}}
  \Tr\qty[ \E^{-\beta H} O ]
  \nonumber\\
  &=
  \frac{1}{\mathcal{Z}}
  \sum_\alpha
  \E^{\beta\mu \sum_i n^f_{\alpha\,i}}
  \det \qty[ \Id + \E^{-\beta \underline H_\alpha} ]
  \frac{\Tr_c \qty[ \E^{-\beta H_\alpha} O ]}{\Tr_c \E^{-\beta H_\alpha}}
  \nonumber\\
  &=
  \sum_\alpha p_\alpha \ev{O}_\alpha,
\end{align}
where
$p_\alpha = P_\alpha / \mathcal{Z}$
is a normalized probability density,
and
$\ev{O}_\alpha$
is the expectation value
for the $f$~electron configuration $\alpha$.

With the above weights
we perform Marcov chain Monte Carlo sampling
on the $f$~electron configurations.
\cite{gubernatis_quantum_2016}
For a given $f$~electron configuration
we use exact diagonalization
to treat the remaining $c$~electron problem.
The expectation value of the observable $O$ is expressed as
\begin{align}
  \ev{O}_\alpha
  &=
  \sum_\nu n_F(\epsilon_\nu) \ev{O}{\nu}
  ,
\end{align}
where $\ket{\nu}$ is an eigenstate of $\underline H_\alpha$
with eigenvalue $\epsilon_\nu$ and
$n_F(\epsilon) = [\exp (\beta \epsilon) + 1]^{-1}$ is the Fermi function.

\subsection{Nonequilibrium}
\label{subsec:nonequilibrium}

To study nonequilibrium phenomena
we apply an instantaneous perturbation to the system at $t=0$.
In particular,
we consider a global interaction quench
\begin{align}
  H(t)
  =
  H
  +
  \sum_i \Delta U(t) \, \cnt{c_i} \cnt{f_i}
  ,
  \label{eq:quench1}
\end{align}
where
$H$ is the equilibrium Hamiltonian as defined in \cref{eq:fkm_hamiltonian}
and
\begin{equation}
  \Delta U(t)
  =
  \theta(t) ( U_q - U )
  \label{eq:quench2}
\end{equation}
is the time-dependent interaction parameter,
defined in terms of the Heaviside function $\theta(t)$.

For ease of presentation
we will refer to the equilibrium Hamiltonian as $H^-$
and to the post-quench Hamiltonian as $H^+$.
The lattice Monte Carlo method described above
extends to the nonequilibrium case.
For each $f$~electron configuration
we diagonalize the $c$~electron Hamiltonians $H^\pm_\alpha$.
The nonequilibrium Hamiltonian $H^+_\alpha$ is time-independent
so that the time-propagation can be performed analytically.

For example the lesser and greater components
of the nonequilibrium Green's functions
between sites $i$ and $j$
for a fixed $f$~electron configuration $\alpha$
are given by
\begin{align}
  {G_\alpha^<}_{ij}(t,t')
  &=
  \I
  \ev*{\cre{c_j}(t') \ann{c_i}(t)}_\alpha
  \nonumber\\
  &=
  \sum_{kl}
  {\underline{u}_\alpha}_{ik}(t)
  \,{G_\alpha^<}_{kl}\,
  {\underline{u}_\alpha^*}_{jl}(t')
  \\
  {G_\alpha^>}_{ij}(t,t')
  &=
  - \I
  \ev*{\ann{c_i}(t) \cre{c_j}(t')}_\alpha
  \nonumber\\
  &=
  \sum_{kl}
  {\underline{u}_\alpha}_{ik}(t)
  \,{G_\alpha^>}_{kl}\,
  {\underline{u}_\alpha^*}_{jl}(t')
  ,
\end{align}
where
the time-propagators are defined as
\begin{equation}
  \underline{u}_\alpha(t)
  =
  \E^{\I t \underline{H}^+_\alpha}
  \qquad
  \underline{u}_\alpha^*(t)
  =
  \E^{- \I t \underline{H}^+_\alpha},
\end{equation}
and the equal-time equilibrium Green's functions are given by
\begin{align}
  - \I {G_\alpha^<}_{ij}
  &=
  \sum_{\nu^-} n_F(\epsilon_{\nu^-}) \ip*{j}{\nu^-}\!\!\ip*{\nu^-}{i}
  \\
  \I {G_\alpha^>}_{ij}
  &=
  \delta_{ij}
  +
  \I {G_\alpha^<}_{ij},
\end{align}
where $\ket{\nu^-}$ are the eigenstates of $\underline{H}^-_\alpha$
and $\epsilon_{\nu^-}$ are the corresponding eigenvalues.

\subsection{Local and Nonlocal Correlation Functions}
\label{subsec:correlations}

The lattice Monte Carlo procedure
gives us access to various local and nonlocal observables
in nonequilibrium.
In particular we study the local $c$~electron density $\rho_i(t)$
and double occupancy $D_i(t)$,
which are given by
\begin{align}
  \rho_i(t)
  &=
  \ev*{\cre{c_i}(t) \ann{c_i}(t)}
  ,
  \\
  D_i(t)
  &=
  \ev*{\cre{c_i}(t) \ann{c_i}(t) \, N^f_i}.
\end{align}
We also study two-point density-density correlations
between $c$~electrons on the sites $0$ and $i$,
and at times $0$ and $t$,
\begin{equation}
  \label{eq:density_density_correlations}
  C_i(t)
  =
  \ev*{n^c_i(t) n^c_0(0)} - \ev*{n^c_i(t)}\!\!\ev*{n^c_0(0)}.
\end{equation}
For a fixed $f$~electron configuration $\alpha$
we can apply Wick's theorem
to evaluate the correlation function
in terms of the single-particle Green's function
\begin{align}
  \label{eq:two_point_density_density}
  \ev*{n^c_i(t) n^c_0(0)}_\alpha
  &=
  {G^<_\alpha}_{0i}(0,t)
  {G^>_\alpha}_{i0}(t,0)
  \\ \nonumber
  &\phantom{=}
  -
  {G^<_\alpha}_{ii}(t,t)
  {G^<_\alpha}_{00}(0,0).
\end{align}

Furthermore, we study the charge susceptibility,
which can be defined as the Fourier transform
of the commutator of the density-density correlations
in both space and time
\begin{align}
  \label{eq:charge_susceptibility}
  \chi_k(\omega)
  &=
  \frac{-\I}{N}
  \sum_r \E^{-\I k r}
  \int_0^\infty \dd{t} \E^{\I \omega t}
  \ev*{[n_r(t), n_0(0)]}.
\end{align}
Inserting
\begin{align}
  \ev*{[n_r(t), n_0(0)]}_\alpha
  &=
  {G^<_\alpha}_{0r}(0,t)
  {G^>_\alpha}_{r0}(t,0)
  \\ \nonumber
  &\phantom{=}
  -
  {G^<_\alpha}_{r0}(t,0)
  {G^>_\alpha}_{0r}(0,t)
\end{align}
and performing the Fourier transform analytically yields
\begin{align}
  \chi_k(\omega)
  &=
  \sum_{ij} \sum_{\mu\nu}
  {G^<_\alpha}_{0i}
  \ip{i}{\mu}
  \kappa_{\mu\nu}
  \mathcal{L}_{\mu\nu}(\omega)
  \ip{\nu}{j}
  {G^>_\alpha}_{j0}
  \\ \nonumber
  &\phantom{=}
  -
  \sum_{ij} \sum_{\mu\nu}
  {G^<_\alpha}_{i0}
  \ip{i}{\mu}
  \kappa_{\mu\nu}
  \mathcal{L}_{\nu\mu}(\omega)
  \ip{\nu}{j}
  {G^>_\alpha}_{0j},
\end{align}
where
\begin{equation}
  \kappa_{\mu\nu}
  =
  \frac{1}{N} \sum_r \E^{-\I k r}
  \ip*{\mu}{r}\!\!\ip*{r}{\nu}
\end{equation}
transforms to reciprocal space, and
\begin{equation}
  \mathcal{L}_{\mu\nu}(\omega)
  =
  \frac{\I}{\omega + \epsilon_\mu - \epsilon_\nu + \I 0^+}.
\end{equation}
We only define the charge susceptibility in equilibrium,
which is why we do not distinguish the quenched and equilibrium Hamiltonian
in this case.

\subsection{Spectral function and optical conductivity}
\label{subsec:spectral_function}

The $c$~electron spectral function
can be obtained by a sampling over the eigenvalues $\epsilon_{\alpha\nu}$
of the $c$~electron Hamiltonian matrix $\underline H_\alpha$
\begin{equation}
  A(\omega)
  =
  \frac{1}{\Z}
  \sum_\alpha p_\alpha
  \sum_\nu \delta(\omega - \epsilon_{\alpha\nu}),
\end{equation}
where we replace the $\delta$-function
by a finite width Lorentzian.
For the zero frequency value it is desirable to avoid Lorentzian broadening
in order to reliably identify a gap opening.
In this case we can take a histogram over all occurring energy eigenvalues
in a finite window around $\omega=0$.

Furthermore, we study the equilibrium optical conductivity,
as previously described by Antipov~\etal,
\cite{antipov_interaction-tuned_2016}
obtained by linear response of the current
to an applied infinitesimal electric field.
Appendix~\ref{subsec:time_dependent_optical_conductivity}
contains a derivation of the more general time-dependent optical conductivity.

\subsection{Dynamical Cluster Approximation}
\label{subsec:dynamical_cluster_approximation}

In Ref.~\onlinecite{herrmann_nonequilibrium_2016}
we studied local and nonlocal correlations
of the Falicov-Kimball model in nonequilibrium after an interaction quench
using the dynamical cluster approximation (DCA).
\cite{maier_quantum_2005}
Since we are interested in comparing the exact (\ie\ converged in lattice size)
results obtained using the above Monte Carlo method
to results from that previous study,
we briefly review the implementation of the nonequilibrium DCA formalism here.

DCA is a cluster extension of the dynamical mean field theory (DMFT).
\cite{hettler_nonlocal_1998,hettler_dynamical_2000,aoki_nonequilibrium_2014}
A cluster of sites $\rcl$ on the lattice is selected
and defines an impurity model
with a self-consistently determined bath.
Additionally, translation invariance and periodic boundary conditions
are imposed on the cluster.
In reciprocal space this yields patches
of reciprocal vectors $\kcl$ of the super lattice
around reciprocal vectors $\Kcl$ of the cluster.
The corresponding effective impurity Hamiltonian reads
\begin{flalign}
  H_\cl  - \mu N_\cl
  &=
  H_0 + H_f
  +
  H_{\mathrm{int}}
  +
  H_{\mathrm{hyb}}
  +
  H_{\mathrm{bath}}
  ,
    \phantom{\Bigg)}
  \\
  H_0
  &=
  \sum_\Kcl \epscl_\Kcl \cre{c_\Kcl} \ann{c_\Kcl}
  -
  \mu \sum_\Kcl \cre{c_\Kcl} \ann{c_\Kcl}
  ,
  \\
  H_f
  &=
  -
  \bigg( \frac{U}{2} + \mu \bigg)
  \sum_{\rcl} \cre{f_{\rcl}} \ann{f_{\rcl}}
  ,
  \\
  H_{\mathrm{int}}
  &=
  \frac{U}{\Ncl}
  \sum_{\Kcl,\Kcl',\rcl} \cre{c_\Kcl} \ann{c_{\Kcl'}}
  \bigg( \cre{f_{\rcl}} \ann{f_{\rcl}} - \frac{1}{2} \bigg)
  e^{-\I(\Kcl-\Kcl')\rcl}
  ,
  \\
  H_{\mathrm{hyb}}
  &=
  \sum_{\Kcl,{\bf p}}
  \big(
    V_{\Kcl,{\bf p}} \cre{c_\Kcl} \ann{a_{\Kcl,{\bf p}}} + \mathrm{h.c.}
  \big)
  ,
  \\
  H_{\mathrm{bath}}
  &=
  \sum_{\Kcl,{\bf p}}
  \varepsilon_{\Kcl,{\bf p}} a^\dagger_{\Kcl,{\bf p}} a_{\Kcl,{\bf p}}
  ,
\end{flalign}
where
$\mu$ is the chemical potential
(here $\mu=0$ corresponds to half-filling),
$\Ncl$ is the number of cluster sites,
$\Kcl$ are the reciprocal cluster vectors,
$\epscl$ is the coarse grained dispersion,
$\rcl$ are the cluster vectors,
$V_{\Kcl,\vb{p}}$ is the hybridization amplitude,
$\varepsilon_{\Kcl,\vb{p}}$ are the bath energy levels,
and $a^{(\dagger)}$ are the bath annihilation (creation) operators.

The impurity problem is solved by explicit summation
over all the $f$~electron configurations.
For each such configuration $\alpha$
we obtain a $c$~electron cluster Green's function $R_{\alpha\Kcl\Kcl'}$.
The final impurity Green's function
is then given by the weighted sum over all $f$~electron configurations
\begin{equation}
  G_\Kcl
  =
  \sum_\alpha
  w_\alpha
  R_{\alpha\Kcl\Kcl}
  ,
\end{equation}
where the
$w_\alpha$ are the equilibrium weights of the $f$~electron configuration.

The self-energy is approximated as constant on each patch in reciprocal space,
$\Sigma_{\Kcl+\kcl} = \Sigma_\Kcl$,
and the self-consistency condition requires the cluster Green's function
to be equal to the coarse grained lattice Green's function
\begin{equation}
  \bar{G}_\Kcl
  =
  \frac{\Ncl}{N} \sum_\kcl G_{\Kcl+\kcl}.
\end{equation}
The nonequilibrium problem is then solved on the Kadanoff-Baym contour.
\cite{herrmann_nonequilibrium_2016}

We can determine the time-dependent double occupancy
straight-forwardly on the impurity model.
By applying Wick's theorem in a similar form as shown above on the cluster impurity
we can also measure nonlocal density-density correlations
\begin{align}
  \langle \cre{c_\rcl} \ann{c_\rcl} \cre{c_{\rcl'}} \ann{c_{\rcl'}} \rangle
  &=
  \sum_\alpha
  w_\alpha
  \langle
    \cre{c_\rcl} \ann{c_\rcl} \cre{c_{\rcl'}} \ann{c_{\rcl'}}
  \rangle_\alpha
  \\ \nonumber
  &=
  \sum_\alpha
  w_\alpha
  \big[
    R^<_{\alpha\rcl'\rcl}
    R^>_{\alpha\rcl\rcl'}
    -
    R^<_{\alpha\rcl\rcl}
    R^<_{\alpha\rcl'\rcl'}
  \big],
  \\
  \ev*{\cre{c_\rcl} \ann{c_\rcl} \cre{f_\rcl} \ann{f_\rcl}}
  &=
  - \I
  \sum_\alpha
  w_\alpha R^<_{\alpha\rcl\rcl} {N^f_\alpha}_{\rcl'}.
\end{align}
It is important to note that this is an approximation
and does not equal the corresponding observable on the whole lattice.
However, for larger cluster sizes these correlations
will converge towards the corresponding lattice observable.

\begin{figure}
  \includegraphics[width=\linewidth]{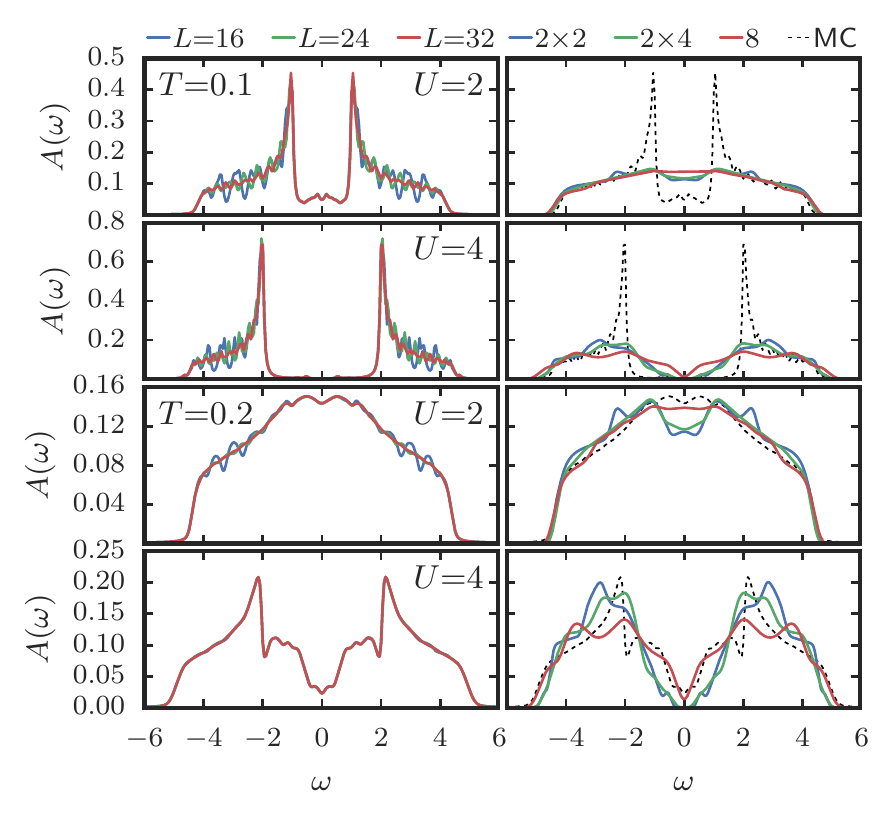}%
  \caption{\label{fig:compare_dca_spectra}%
    Convergence of the equilibrium spectral function
    with respect to the linear cluster size $L$
    in the lattice Monte Carlo simulations
    (left column).
    Comparison between
    DCA calculations
    at different DCA cluster geometries
    without averaging over patch layouts
    and lattice Monte Carlo results for $L=32$
    (right column).
  }
\end{figure}

\begin{figure*}
  \includegraphics[width=\linewidth]{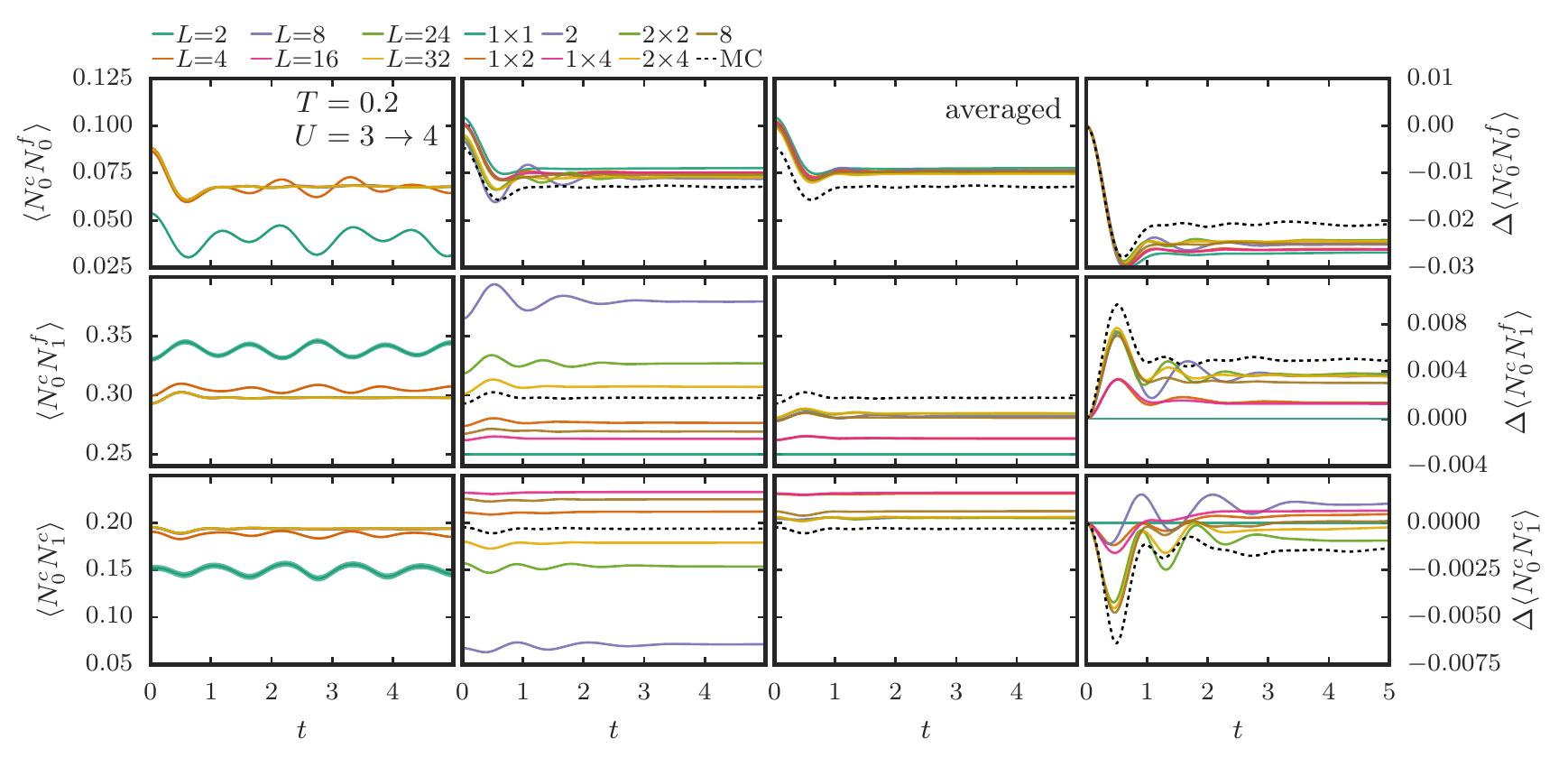}%
  \caption{\label{fig:compare_dca}%
    Two-dimensional Falicov-Kimball model: Convergence of double occupancy,
    $c$-$f$, and $c$-$c$~nearest-neighbor density-density correlations
    as a function of time
    after an instantaneous global interaction quench
    from $U_0=3$ to $U_q=4$ at temperature $T=0.2$
    with respect to the linear cluster size $L$
    in the lattice Monte Carlo simulations
    (first column).
    Comparison between
    DCA calculations
    for different DCA cluster geometries
    and lattice Monte Carlo results at $L=32$
    after the same interaction quench
    (second column).
    Comparison between
    DCA calculations
    averaged over patch layouts
    as presented in Ref.~\onlinecite{herrmann_nonequilibrium_2016}
    and lattice Monte Carlo results at $L=32$
    after the same interaction quench
    (third column).
    Time evolution of local and nonlocal correlation functions
    relative to their initial value
    ($y(t) - y(t_0)$)
    (fourth column).
  }
\end{figure*}

\section{Results}

\subsection{Convergence of lattice Monte Carlo and DCA}

We start by demonstrating the convergence of the lattice Monte Carlo results
with increasing lattice size,
and then use the converged Monte Carlo results
to benchmark the DCA calculations.
We consider the two-dimensional model with nearest neighbor hopping,
and use this nearest-neighbor hopping as the unit of energy.

\subsubsection{Equilibrium}

Figure~\ref{fig:compare_dca_spectra} shows equilibrium spectral functions
for $U=2,4$ and temperatures $T=0.1,0.2$.
The left panels show the lattice Monte Carlo results
for indicated linear size $L$ of the cluster,
while the right hand panels compare different DCA spectra
to the lattice Monte Carlo result for $L=32$.
At the higher temperature the system is in the disordered phase
\cite{antipov_interaction-tuned_2016} for all interactions
(see \cref{fig:phase_diagram}),
and the Monte Carlo spectra are essentially converged for $L=24$.
At $U=4$, the system is in the crossover region
to the Anderson insulator regime,
and a pseudo-gap opens in the spectral function.
The comparison with the DCA spectra (right panels)
shows an overall good agreement already for small cluster size.
While the 2 $\times$ 2  and $2\times 8$ clusters overestimate the pseudo-gap
due to strong charge-order tendencies,
the diagonal 8-site cluster underestimates it.

At the lower temperature $T=0.1$,
the infinite lattice system is in a charge-ordered insulating phase
(see \cref{fig:phase_diagram}),
with the $U=2$ case close to the phase boundary to the Anderson insulator.
Accordingly, there is a large gap in the spectral function.
As demonstrated in the left panels of \cref{fig:compare_dca_spectra}
the convergence of the Monte Carlo spectra with cluster size is slower,
due to spiky features.
Nevertheless it is clear that the $L=32$ result
is up to small oscillations converged.
The DCA results qualitatively differ from the lattice Monte Carlo spectra,
and rather resemble the high-temperature results.
This is because of the translation invariance which is enforced on the cluster,
and the suppression of long-range order in the self-consistency.
The comparison between the DCA and lattice Monte Carlo results
shows how the appearance of strong charge order correlations
opens a gap of size $U$
and shift the spectral weight from the gap region
to sharp peaks at the gap edge.

\subsubsection{Nonequilibrium}

We next consider the time evolution after an interaction quench.
To reduce discrepancies originating from charge order, we set $T=0.2$.
Figure~\ref{fig:compare_dca}
shows the simulation results for a quench from $U=3$ to $U=4$.
The left panels illustrate the convergence of the lattice Monte Carlo results
with the linear lattice size $L$
(the width of the curves corresponds to the Monte Carlo error).
Looking at the initital value,
one sees that the result for the local observable $\langle N^c_0 N^f_0\rangle$
converges very rapidly with lattice size,
and that even the nonlocal quantities
$\langle N^c_0 N^f_1\rangle$ and $\langle N^c_0 N^c_1\rangle$
are converged already for $L=8$.
While the time evolution for $L=2$ and $L=4$ exhibits spurious oscillations,
for $L \ge 8$ the dynamics shows a rapid damping
and the curves are converged up to time $t=5$.
We can thus use these converged Monte Carlo data
to benchmark the quench dynamics predicted by nonequilibrium DCA.

The second row compares the DCA evolution for different cluster geometries
to the exact Monte Carlo result (black dashed curve).
We first of all note that for the moderate cluster sizes considered,
there is a strong cluster size dependence in the DCA results,
especially for the nonlocal observables.
While the results for the largest clusters tend to be relatively close
to the benchmark curve,
and the damping behavior is qualitatively well reproduced
for the larger clusters,
there is no systematic convergence with cluster size.
The situation can be improved by calculating averages
over different cluster geometries (patch layouts),
as discussed in Ref.~\onlinecite{herrmann_nonequilibrium_2016}.
In the third column,
we compare these averaged curves to the lattice Monte Carlo result.
One now obtains a systematic convergence towards the exact result
with increasing lattice size,
although the off-set to the Monte Carlo curve
remains substantial for cluster size 8
and in particular larger than what one might have guessed
based on the difference between the ``$4 \times 2$" and ``$8$" simulations.
These deviations,
which are also evident in the equilibrium spectra
plotted in \cref{fig:compare_dca_spectra},
may have several origins:
the suppression of charge-order correlations in DCA,
the lack of vertex corrections outside the cluster,
or the fact that Anderson insulator behavior
cannot be captured on small-size clusters
(the $U$ value after the quench
is at the border of the Anderson insulator regime).

To better judge the accuracy of the damping behavior,
we remove the offsets by subtracting the value at time $t=0$
from all the curves.
The corresponding results are shown in the right panels.
We see that the damping dynamics is qualitatively well reproduced
by the larger clusters,
but differences remain at longer times and,
for the nonlocal observables,
even the initial response to the quench is not quantitatively accurate.
While the absolute changes of the correlations are small,
our results indicate that clusters with substantially more than 8 sites
are needed to fully converge the DCA calculations.
Comparing the 8-site DCA results
to the $4 \times 4$ sites lattice Monte Carlo curve,
we conclude that the DCA construction
speeds up the convergence of the damping behavior,
but results in a significant offset
of the local and nonlocal correlation functions,
so that the convergence to the exact infinite-lattice result
is faster in the lattice Monte Carlo approach.

\subsection{Spreading of Correlations}

\subsubsection{One-dimensional model}

\begin{figure*}
  \includegraphics[width=\linewidth]{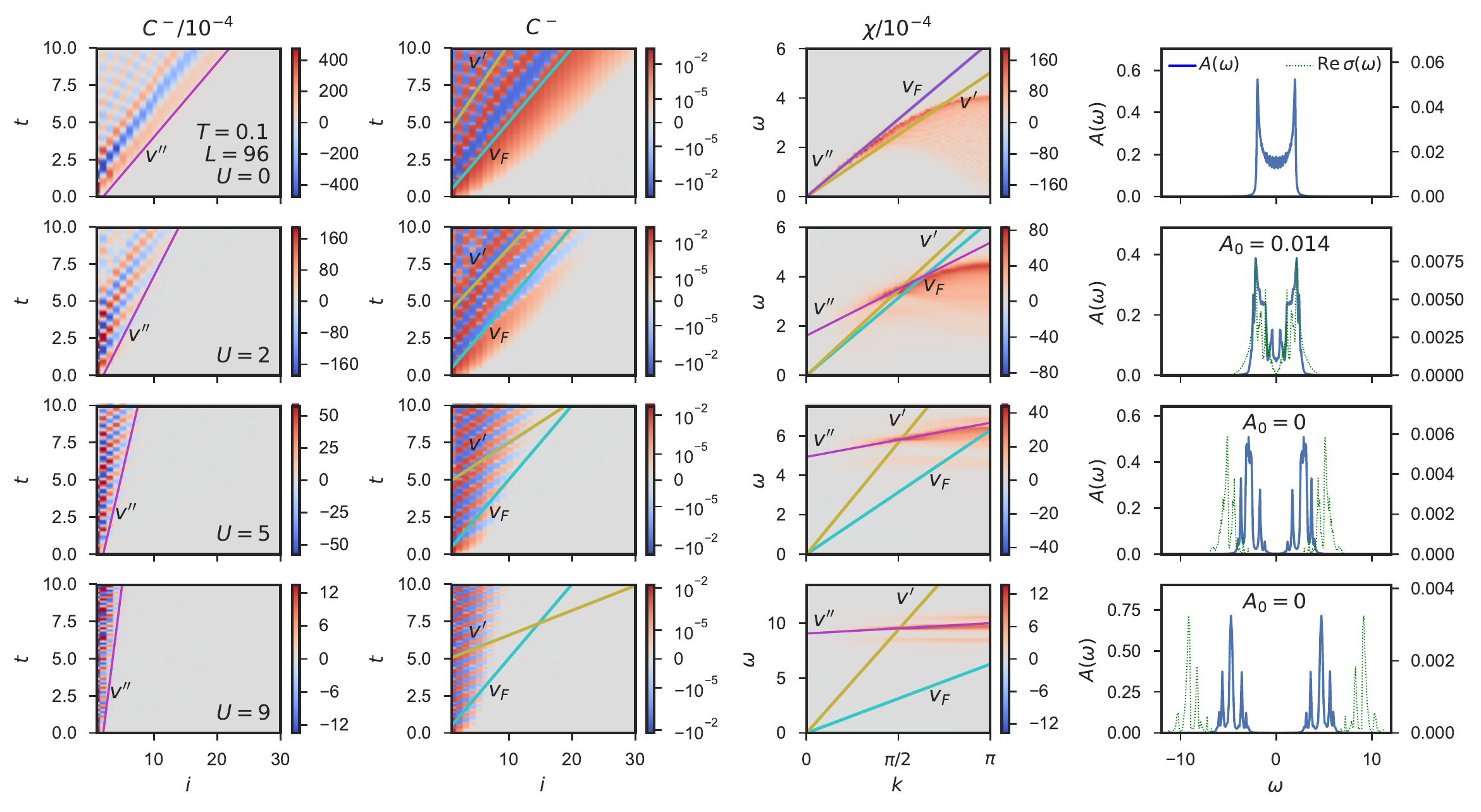}\\
  \includegraphics[width=\linewidth]{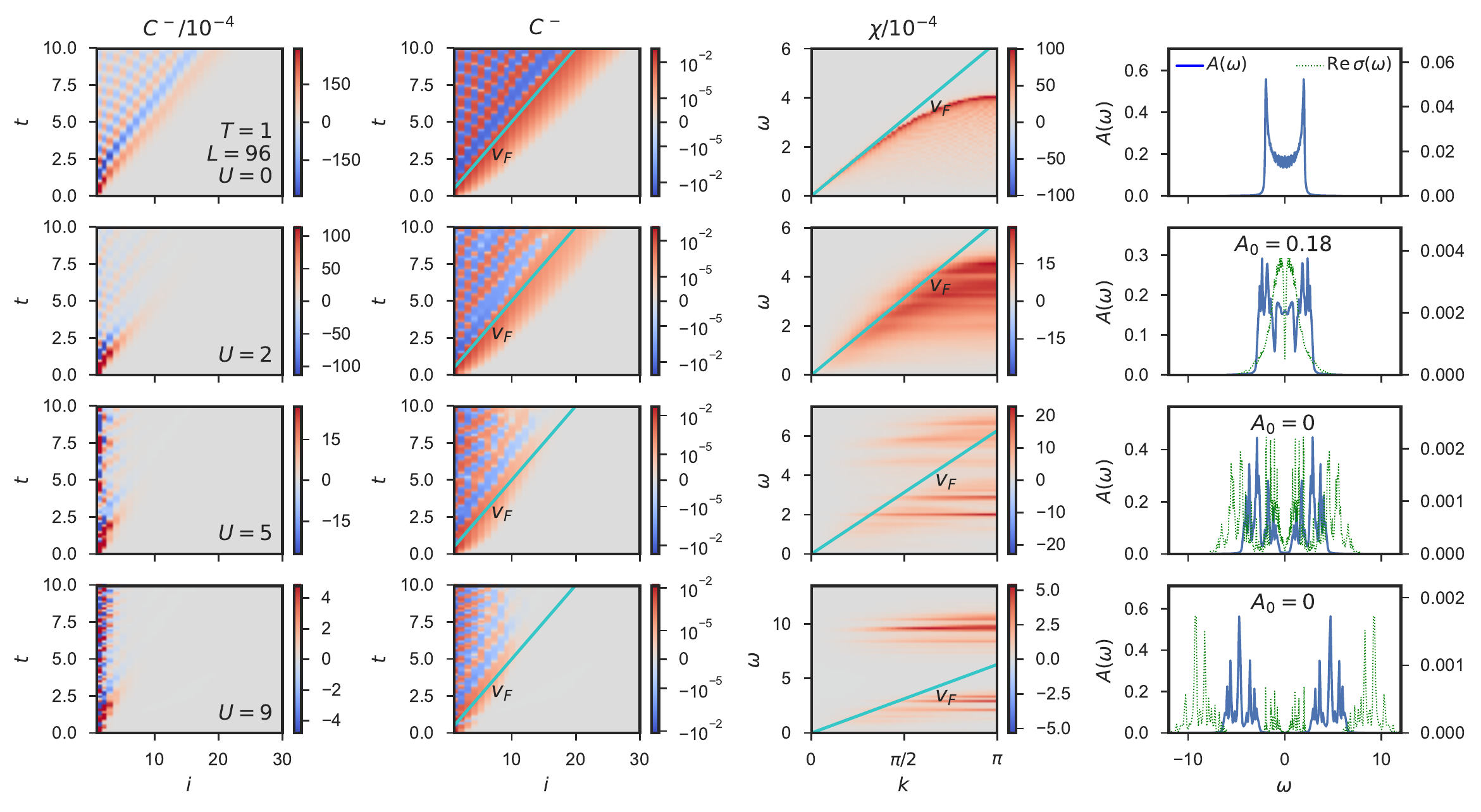}%
  \caption{\label{fig:spreading_correlations_1d}%
    One-dimensional model:
    Spreading of two-point density-density correlations
    as defined in
    \cref{eq:density_density_correlations,eq:spreading_correlations}
    in an equilibrium system of size $L=96$
    at temperatures $T=0.1$ and $T=1$
    for various interaction parameters.
    The first column shows
    $C^-_i(t)$ on a linear scale,
    while the second column shows the same data on a logarithmic scale.
    The third column plots the charge susceptibility
    as a function of frequency $\omega$ and reciprocal vector $k$.
    The fourth column shows the local density of states
    and the real part of the optical conductivity
    at the given parameters.
    These curves are subject to Lorentz-broadening.
    The zero frequency value $A_0$ is determined separately
    as described in \cref{subsec:spectral_function}
    in order to avoid the effects of Lorentz broadening.
  }
\end{figure*}

We start by discussing the properties
of the equilibrium one-dimensional Falicov-Kimball model,
where lattices up to $L=128$ can be treated.
Here we expect metallic, WL/AI and MI behavior with increasing $U$,
even at low temperature.
The simulation results are presented in \cref{fig:spreading_correlations_1d}
for different interactions $U$
and two values of the temperature, $T = 0.1$ and $1$.

In the right hand panels, we plot the density of states
and the real part of the optical conductivity.
The $U=2$ system has a finite density of states at $\omega=0$,
but a vanishing conductivity.
According to the analysis of Ref.~\onlinecite{antipov_interaction-tuned_2016},
this indicates an AI or WL state.
(To distinguish the two,
one would have to study the scaling of the conductivity with system size.)
The states at $U = 5$ and $9$ are
characterized by a vanishing density of states and correspond to the MI phase.

To study the spreading of the density-density correlations with time
we calculate the commutator
\begin{equation}
  \label{eq:spreading_correlations}
  C^-_i(t)
  =
  \ev*{[n_i(t), n_0(0)]}
  =
  2 \I \Im[ \ev*{n^c_i(t) n^c_0(0)} ].
\end{equation}
The left two columns of \cref{fig:spreading_correlations_1d}
show the results on a linear and logarithmic scale.
The top four panels correspond to low temperature ($T=0.1$)
and the bottom four panels to high temperature ($T=1$).
At $U=0$, the correlations spread with the Fermi velocity $v_F=2$
up to distances comparable to the system size.

For $U>0$,
the spreading of correlations is not linear in time any more.
The correlations only extend up to some maximum distance,
which decreases with increasing interaction strength,
indicative of localization behavior.
The localization length is the smallest in the MI phase.
This behavior is in contrast to the Mott phase of the Hubbard model,
which is characterized by freely propagating spin modes.
In the Falicov-Kimball model,
the spreading of the charge excitations
is quenched in the disordered MI, in a way analogous to the AI phase.

We also notice that the spreading velocity
is reduced with increasing interactions,
which is most clearly visible
in the low-temperature data plotted on the linear scale.
To determine the spreading velocity,
we look at the charge susceptibility $\chi_k(\omega)$
defined in \cref{eq:charge_susceptibility},
which is plotted in the third row.
The maximum slope in the $\omega$-versus-$k$ curves
defines the spreading (group) velocity $v''$
and the corresponding light cone is plotted (with an arbitrary offset)
in the left-hand panels.
While $v''$ can be rather easily determined
in the low-temperature simulation results,
the $\chi_k(\omega)$ plots for $T=1$
show both dispersing and flat features,
so that the definition of $v''$ becomes ambiguous.
One noteworthy point is that in the high-temperature system,
the Fermi velocity of the noninteracting model
controls the spreading of correlations at short times,
even in the AI and MI regimes.

In contrast to the group velocity, the phase velocity $v'$,
\ie\ the velocity of the wave fronts,
increases with increasing interaction strength.
While there is a small speed-up of the phase velocity with time,
it can be unambiguously determined at a given time in the low-temperature data.
We extracted $v'$ by fitting the wave front emerging from $t \approx 5$
as shown by the yellow line in the second row.
If one indicates the corresponding velocity in the plot of $\chi_k(\omega)$,
it matches the maximum intensity point in the susceptibility.
At $T=1$, the definition of $v'$ is more difficult,
since at early times, the phase velocity is essentially given by $v_F$,
while at some later time,
an enhanced phase velocity reminiscent of the low-temperature data appears,
at least at large $U$.
In the $\chi_k(\omega)$ plot, there are correspondingly two branches --
the upper branch looks similar to the low-temperature susceptibility,
while the lower branch resembles the noninteracting dispersion.
The large $v'$ roughly explains the edge of the upper branch.

The top panel of \cref{fig:velocities}
shows the $U$-dependence of the velocities $v'$ and $v''$
for the one-dimensional model at $T=0.1$.
For $U \gtrsim 2$, \ie\ in the MI regime,
the phase velocity $v'$ scales linearly with $U$,
while the spreading velocity $v''$ scales roughly like $1/U$.

\subsubsection{Two-dimensional model}

\begin{figure*}
  \includegraphics[width=\linewidth]{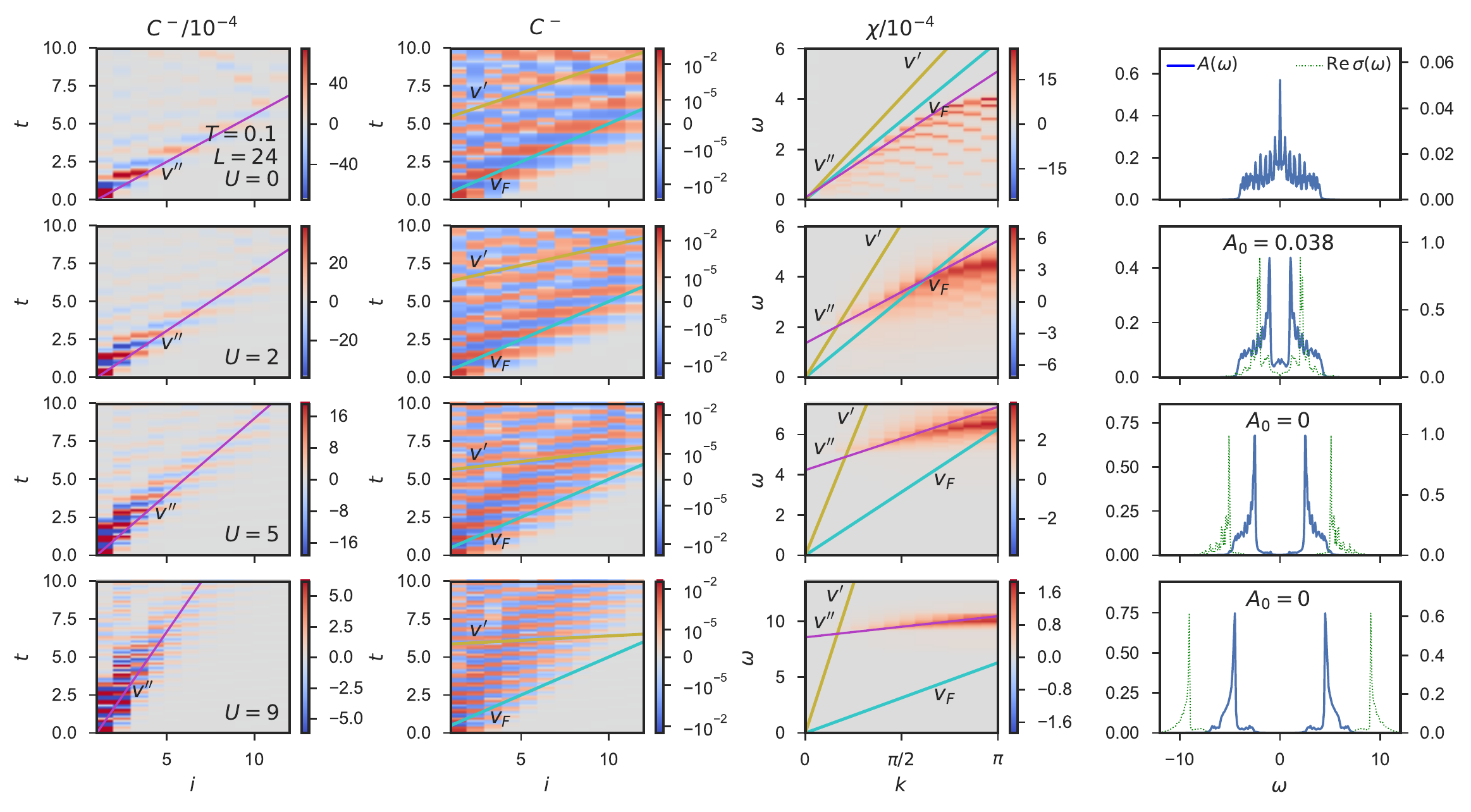}\\
  \includegraphics[width=\linewidth]{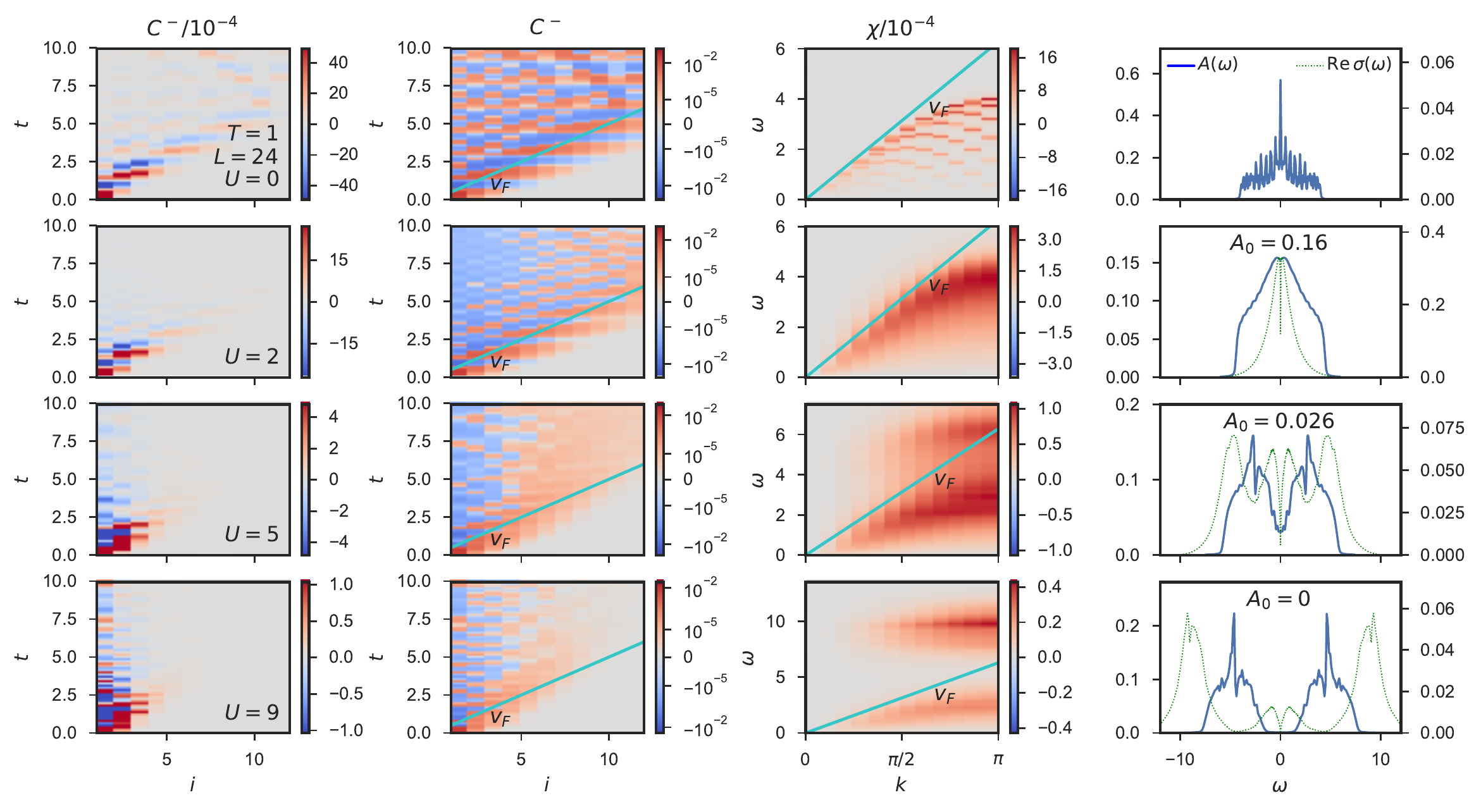}%
  \caption{\label{fig:spreading_correlations}%
    Two-dimensional model:
    Spreading of two-point density-density correlations
    as defined in
    \cref{eq:density_density_correlations,eq:spreading_correlations}
    in an equilibrium system of linear size $L=24$
    at temperatures $T=0.1$ and $T=1$
    for various interaction parameters.
    The first column shows
    $C^-_i(t)$ on a linear scale,
    while the second column shows the same data on a logarithmic scale.
    The third column plots the charge susceptibility
    as a function of frequency $\omega$ and reciprocal vector $k$.
    The fourth column shows the local density of states
    and the real part of the optical conductivity
    at the given parameters.
    These curves are subject to Lorentz-broadening.
    The zero frequency value $A_0$ is determined separately
    as described in \cref{subsec:spectral_function}
    in order to avoid the effects of Lorentz broadening.
  }
\end{figure*}

We now turn to the two-dimensional model,
where simulations up to linear size $L=24$ are possible with modest resources.
The results for $T=0.1$ and $1$ are shown in \cref{fig:spreading_correlations}.
At the higher temperature,
the $U=0$, $2$, $5$, and $9$ panels
correspond to the M, WL, AI, and MI regimes,
as evidenced by the density of states and conductivity data
(see also \cref{fig:phase_diagram}).
At the lower temperature,
$U=2$ is at the border between WL and CDW,
while $U=5$ and $9$ are in the CDW phase.
In the metallic phase ($U=0$),
the correlations spread at the fastest Fermi velocity in the $x$-direction,
which is $v_F=2$.
(Because of the small system size,
the left and right wave fronts cross around $t \approx 6$,
which is evident at later times.)
For $U>0$, the low-temperature data reveal a reduced spreading velocity $v''$,
but up to time $t=10$,
there is no evidence for a finite localization length.
The spreading behavior in the CDW phase
is thus similar to the expected result for the Hubbard model.
At $T=1$, on the other hand,
one observes a localization behavior in the WL/AI and disordered MI phases.
Compared to the one-dimensional case,
the phase velocity $v'$ is substantially larger,
and near the edge of the light cone almost independent of $U$.
However, at shorter distances,
one can also identify slower phase velocities,
as indicated by the yellow lines in the second column.

\begin{figure}
  \includegraphics[width=\linewidth]{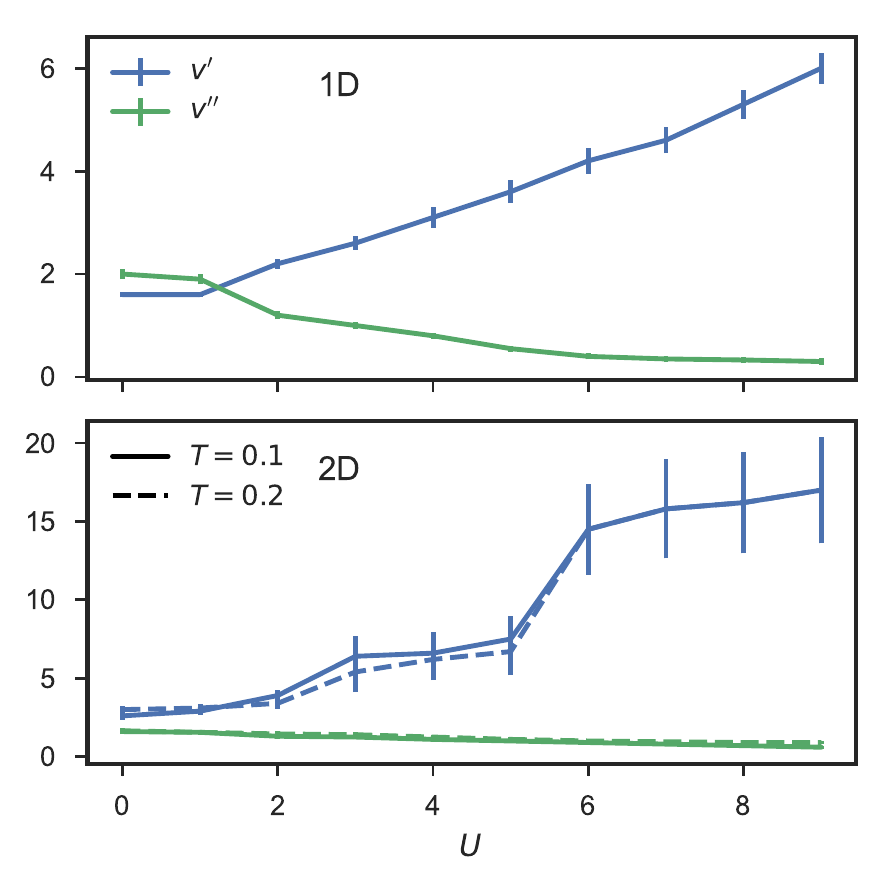}%
  \caption{\label{fig:velocities}%
    Phase velocity $v'$ and spreading velocity $v''$
    as a function of the interaction parameter $U$
    in a one-dimensional model with $L=96$ (upper panel)
    and a two-dimensional model with $L=24$ (lower panel)
    at temperature $T=0.1$ (solid line) and $T=0.2$ (dashed line).
    The velocities were fitted as shown in
    \cref{fig:spreading_correlations_1d,fig:spreading_correlations},
    the errorbars indicate the estimated accuracy.
  }
\end{figure}

We plot the $U$-dependence of this slower phase velocity $v'$
together with the light cone velocity $v''$
in the bottom panel of \cref{fig:velocities}, for $T=0.1$ and $0.2$.
The phase velocity shows a jump between $U=5$ and $6$,
which can be associated with the AI to MI transition at the higher temperature.
Interestingly, the same feature persist even at $T=0.1$,
in the CDW insulating phase,
which has long-range ordered $f$~particle cofigurations.
This indicates that the phase velocity
is influenced more by correlation effects than by disorder effects.
(Note that correlation induced changes near the Mott transition value of $U$
occur also inside the CDW phase,
similar to the crossover from weak-coupling to strong-coupling antiferromagnet
in the Hubbard model.\cite{fratino_signatures_2017})

In the high-temperature system ($T=1$),
the spreading at short times is controlled by the $v_F$
of the noninteracting system, even for large $U$.
This is the same behavior as already observed in the one-dimensional case,
and it is again reflected in the susceptibility
in the form of two branches --
a lower branch with a maximum slope given by $v_F$
and a weakly dispersing upper branch
whose maximum slope is related to the spreading velocity at later times.

We have also studied the spreading of correlations
after an interaction quench between the CDW/MI and WL phases
of the two-dimensional model.
The simulation results show a spreading behavior similar to a ``cold" system
at the final interaction,
even though a substantial amount of energy is injected by the quench.
They also provide further support for our observation that CDW correlations
in the disorder potential enhance the spreading range.
The details are presented in Appendix~\ref{AppendixB}.

\section{Discussion and Conclusions}

\begin{table}[t]
  \centering
  \setlength{\tabcolsep}{12pt}
  \begin{tabular}{ccc}
    \toprule
    & increasing $U$ & increasing CWD \\
    \colrule
    range & $\searrow$ & $\nearrow$\\
    $v''$ & $\searrow$ & $\searrow$ \\
    $v'$ & $\nearrow$ & --- \\
    \botrule
  \end{tabular}
  \caption{\label{tab:trends}%
    Summary of the observed effect of
    increasing interaction parameter $U$, and
    increasing charge-density wave order
    on the interaction range,
    the spreading velocity $v''$, and
    the phase velocity $v'$.
  }
\end{table}

We have studied dynamical properties of the Falicov-Kimball model
using Monte Carlo simulation with exact time propagation.
In particular, we focused on the spreading of density-density correlations
in the half-filled one- and two-dimensional model
and related the observed behavior to the equilibrium phase diagram.
There are three quantities which characterize the spreading:
the maximum range of the correlations,
the spreading (or group) velocity $v''$, and
the phase velocity $v'$.
The value of these quantities depends on the interaction $U$,
which determines the strength of the disorder potential,
and on the CDW correlations in the disorder potential.
We summarize the general trends in \cref{tab:trends}.

In the $U=0$ metallic regime, the correlations are not bounded,
and they spread with the maximum Fermi velocity.
For $U>0$, in the disordered phase, the range of the correlations is limited,
in accordance with Anderson localization.
This holds both for the WL/AI phase at weak and intermediate $U$
and the MI phase at large $U$.
There is a systematic trend of decreasing correlation range with increasing $U$.

On the other hand, in the CDW phase of the two-dimensional model,
we do not observe any indications of localization behavior.
While the accessible system size is limited,
one can clearly conclude
(by comparing the behavior above and below the CDW transition temperature)
that the correlations in the disorder potential
allow the density-density correlations to spread farther.
In fact, in the CDW phase, the correlations spread without apparent bound,
independent of $U$, and by fixing the $f$~particle configuration
to a perfect CDW pattern,
one can furher increase the correlations, compared to the thermal ensemble.
We note that in the thermodynamic limit of the model
\cite{lemanski_gapless_2014}
multiple CDW phases can be present
at different values of the interaction strength and temperature,
and that their dynamical signatures may vary.
In the CDW phase at $T=0.1$ all possible localized states are gapped out.

We next consider the spreading (or light-cone) velocity $v''$.
It is obvious from the simulation data that this velocity
decreases with increasing $U$,
and that in this case, too,
there is no dramatic change at the transition from the WL/AI phase
to the MI phase.
Even within the CDW phase,
there is a systematic trend of decreasing $v''$ with increasing $U$.
In fact, in this case one finds that the effects of $U$ (disorder strength)
and disorder correlations cooperate in reducing $v''$.
This is for example evident
by comparing $v''$ above and below the CDW transition temperature,
or the fact that a simulation with a perfect CDW disorder potential
leads to a slower spreading than the simulation for the thermal ensemble.

As for the phase velocity $v'$
one finds a systematic increase with increasing $U$.
This trend is most evident in the data for the one-dimensional model,
where the wave fronts can be clearly identified.
In the two-dimensional case, the phase velocity at the edge of the light cone
becomes very large.
However, at short distances and later times,
one can also identify ``slow'' phase velocities,
which resemble the behavior in the one-dimensional case.
These $v'$ exhibit a jump at the transition from the WL/AI to the MI phase.
Interestingly, the same jump is found
even below the CDW phase transition temperature,
which indicates that CDW correlations
have no important effect on the phase velocity.

While the above analysis holds for low temperatures,
the spreading behavior at higher temperatures is more complicated.
Here, we find that the initial spreading is determined by the Fermi velocity
of the noninteracting system, while at later times,
it is impossible to measure $v''$ due to strong localization effects.
Also the phase velocity becomes difficult to measure for $U>0$.
The charge susceptibility of the high temperature systems
is characterized by two dispersing features,
one corresponding to the initial spreading with velocity $v_F$,
and the other to localized charge excitations.

We also used the quench set-up to study the convergence properties
of lattice Monte Carlo and DCA simulations.
The Monte Carlo results
converge rapidly with lattice size.
Simulations on an $8 \times 8$ system
are sufficient to produce essentially exact results for times up to $t=5$,
which is enough to observe a complete damping of transient oscillations
in local and nearest-neighbor correlation functions.
The DCA results, on the other hand,
exhibit a very strong dependence on the cluster size and geometry
on the small clusters ($\le 8$ sites) that are accessible with
our implementation.
While averaging over different patch layouts
improves the convergence with cluster size,
substantial deviations from the exact result remain
for these small clusters,
and a reliable extrapolation to the thermodynamic limit is not yet possible.
Given the slow convergence of DCA with cluster size,
it is worthwhile to discuss the computational effort of DCA
compared to direct lattice Monte Carlo simulations.

The DCA implementation used in this work
and in Ref.~\onlinecite{herrmann_nonequilibrium_2016}
solves the Falicov-Kimball model
by exact enumeration of all $f$~electron configurations.
Both the computational effort and required memory for this method
scale exponentially as $2^\Ncl$,
which prevents us from simulating clusters of more than eight sites.
The Monte Carlo method used in this work, on the other hand,
does not suffer from such an exponential scaling.
Experience shows that even a $24 \times 24$ cluster
only requires about $2^{13}$ Monte Carlo measurements.
Additionally, the Monte Carlo procedure is trivially parallelizable,
requiring no synchronization except for the final statistical analysis.
It is possible to also implement DCA
using Monte Carlo sampling over the $f$~particle configurations,
and it would be interesting to study how far such a Monte Carlo based
scheme can be pushed.

Another essential difference is
that the nonequilibrium DCA
employs a time-stepping algorithm
\cite{herrmann_nonequilibrium_2016}
that scales cubically in the number of time steps for computation
and quadratically for memory requirements.
Furthermore, the time discretization cannot be chosen arbitrarily,
but has to be chosen small to ensure a converged solution.
The Monte Carlo method, on the other hand,
scales linearly in the number of time steps,
for both computation and memory,
and the time grid can be chosen arbitrarily
due to the analytic time-propagation.

For comparison, simulating an 8-site cluster using DCA on 16 cores
requires more than $100\,\mathrm{GB}$ of memory
and takes about eight hours,
while a Monte Carlo simulation of an $8 \times 8$ cluster on the same machine
requires about $200\,\mathrm{MB}$ of memory
and takes about ten minutes.
In view of these considerations,
it seems that the cluster DMFT approach
does not offer any particular advantages
in the study of the Falicov-Kimball model,
and that the direct lattice simulation is the better strategy,
even in the nonequilibrium case,
or for the calculation of dynamical response functions.

\acknowledgements

We thank D.~Golez, M.~Foster, Y.~Murakami and S.~Kehrein
for helpful discussions.
AH and PW acknowledge support from ERC starting grant No.~278023.
Some part of this work was performed at the Aspen Center for Physics,
which is supported by National Science Foundation grant PHY-1607761.

\appendix

\section{Time-Dependent Optical Conductivity}
\label{subsec:time_dependent_optical_conductivity}

To study the system's linear response
to a periodic electromagnetic field
we can follow the procedure described
by Maekawa \etal,\cite{maekawa_physics_2004}
extended to nonequilibrium
by Lenar\v{c}i\v{c} \etal\cite{lenarcic_optical_2014}
The effect of an electromagnetic field at time $t'$
is determined by the vector potential $\vb{A}(t')$.
To arrive at a linear response formalism
we expand the Hamiltonian to second order:
\begin{align}
  H_e \qty\big(\vb{A}(t'))
  &=
  {-t} \sum_{\nn{ij}}
  \exp\!\qty\big( \I e \vb{A}(t') \cdot \vb{r}_{ij} )\,
  \cre{c_j}\ann{c_i} + H^+_{\mathrm{int}}
  \\
  &\approx
  H^+
  - e \vb{A}(t') \cdot \vb{j}
  + \frac{e^2}{2} \vb{A}(t') \cdot \tau \vb{A}(t')
  \\
  &=
  H^+ + H'(t')
  ,
\end{align}
where $e$ is the elementary charge,
$\vb{r}_{ij} = \vb{r}_j - \vb{r}_i$
is the vector between two sites,
and the current and stress tensor operator
are given by
\begin{align}
  \vb{j}
  &=
  \sum_{ij} \vb{j}_{ij}
  =
  \I t \sum_{\nn{ij}} \vb{r}_{ij} \cre{c_j}\ann{c_i},
  \\
  \tau
  &=
  \sum_{ij} \tau_{ij}
  =
  t \sum_{\nn{ij}}
  \vb{r}_{ij} \cdot \vb{r}_{ij}\tran\,
  \cre{c_j}\ann{c_i}.
\end{align}
We note that
the outer product of $\vb{r}_{ij}$ with itself
is diagonal in the canonical basis
in the case of a two-dimensional square lattice
with nearest neighbor hopping.
The electrical current is then given by
\begin{equation}
  \vb{j}_e(t')
  =
  -\pdv{H\qty\big(\vb{A}(t'))}{\vb{A}(t')}
  =
  e\,\vb{j} - e^2 \tau \vb{A}(t').
\end{equation}

We separate the nonequilibrium Hamiltonian from the perturbation
in the time propagation operator
and expand to first order.
The time propagation from $t$ to $t'$ is then given by
\begin{align}
  U(t', t)
  &=
  \E^{- \I (t'-t) H^+}
  U'(t', t)
  ,
  \\
  U'(t', t)
  &=
  1 - \I \int_t^{t'} \dd{\bar{t}} H'^I(\bar{t})
  ,
\end{align}
where the interaction picture operators are given by
\begin{equation}
  O^I(t)
  =
  \E^{\I t H^+}
  O
  \E^{-\I t H^+}.
\end{equation}

The current expectation value is
\begin{align}
  \ev{\vb{j}_e}(t')
  =
  {}
  &
  e \ev{\vb{j}^I(t')}
  -
  e^2 \ev{\tau^I(t')} \vb{A}(t')
  \\ \nonumber
  {}
  &
  +
  e^2
  \int_t^{t'} \dd{\bar{t}}
  \chi(t',\bar{t})
  \vb{A}(\bar{t})
  ,
\end{align}
where
\begin{equation}
  \chi_{ab}(t', \bar{t})
  =
  \I \theta(t' - \bar{t})
  \ev{[ \vc{j}_a^I(t'), \vc{j}_b^I(\bar{t}) ]}
\end{equation}
is the current-current correlation function
in the real-space components $a,b \in \{x, y, z\}$.

We define the conductivity $\sigma$
as the system's response to the electric field
\begin{equation}
  \delta \ev{\vb{j}_e(t')}
  =
  V
  \int_t^{t'} \dd{t''}
  \sigma(t', t'')\,\delta \vb{E}(t'')
  ,
\end{equation}
where $V$ is the volume.
Taking into account that
\begin{equation}
  \vb{A}(t') = - \int_t^{t'} \dd{t''} \vb{E}(t''),
\end{equation}
one finds
\begin{align}
  \sigma(t', t)
  &=
  \frac{e^2}{V}
  \qty[
    \ev{\tau^I(t')}
    -
    \int_t^{t'} \dd{\bar{t}}
    \chi(t', \bar{t})
  ].
\end{align}
We define the time and frequency dependent conductivity as
\begin{equation}
  \sigma(\omega, t)
  =
  \int_0^\infty \dd{s}
  \sigma(t + s, t)
  \E^{\I \omega s}
  ,
\end{equation}
and separate the Drude weight (or stiffness) $D(t)$
as the dissipationless component such that
\begin{equation}
  \Re \sigma(\omega, t)
  =
  2\pi e^2 D(t) \delta(\omega)
  +
  \Re \sigma_{\mathrm{reg}}(\omega, t).
\end{equation}
When evaluating the Fourier transform
we use the following transform of the Heaviside function
\begin{align}
  \int_{-\infty}^{\infty} \dd{t}
  \E^{\I \omega t}
  \theta(t)
  =
  \frac{%
    \I
  }{%
    \omega + \I 0^+
  }
  =
  \pv \frac{\I}{\omega}
  +
  \pi \delta(\omega)
  ,
\end{align}
where $\pv$ indicates the Cauchy principal value.
Finally,
we arrive at the following expressions
for the time and frequency dependent conductivity and the Drude weight
at a fixed $f$~electron configuration:
\begin{widetext}
\begin{align}
  \sigma_{ab}(\omega, t)
  =
  {}
  &
  \frac{e^2}{V}
  \sum_{mn}
  n_{mn}
  \frac{%
    \I
    \E^{-\I (\epsilon^+_m - \epsilon^+_n) t}
  }{%
    \omega - \epsilon^+_m + \epsilon^+_n + \I 0^+
  }
  \bigg[
    {\tau_{ab}}_{mn}
    +
    \sum_o
    \frac{%
      {j_a}_{no}
      {j_b}_{om}
    }{%
      \omega - \epsilon^+_o + \epsilon^+_n + \I 0^+
    }
    -
    \sum_o
    \frac{%
      {j_b}_{no}
      {j_a}_{om}
    }{%
      \omega - \epsilon^+_m + \epsilon^+_o + \I 0^+
    }
  \bigg]
  ,
  \\
  D_{ab}(t)
  =
  {}
  &
  \frac{1}{2V}
  \sum_{\epsilon^+_m = \epsilon^+_n}
  n_{mn}
  \bigg[
    {\tau_{ab}}_{mn}
    +
    \sum_{\epsilon^+_o \neq \epsilon^+_n}
    \frac{%
      {j_a}_{no}
      {j_b}_{om}
    }{%
      \epsilon^+_n
      -
      \epsilon^+_o
    }
    -
    \sum_{\epsilon^+_o \neq \epsilon^+_m}
    \frac{%
      {j_b}_{no}
      {j_a}_{om}
    }{%
      \epsilon^+_o
      -
      \epsilon^+_m
    }
  \bigg]
  \\ \nonumber
  &
  {}
  +
  \frac{1}{2V}
  \sum_{\epsilon^+_m \neq \epsilon^+_n}
  n_{mn}
  \frac{%
    \E^{-\I (\epsilon^+_m - \epsilon^+_n) t}
  }{%
    \epsilon^+_n - \epsilon^+_m
  }
  \bigg[
    \sum_{\epsilon^+_o = \epsilon^+_n}
    {j_a}_{no}
    {j_b}_{om}
    -
    \sum_{\epsilon^+_o = \epsilon^+_m}
    {j_b}_{no}
    {j_a}_{om}
  \bigg]
  ,
\end{align}
\end{widetext}
where
\begin{align}
  \tau_{nm}
  &=
  \tau_{mn}
  =
  t
  \sum_{\nn{ij}}
  \vb{r}_{ij} \cdot \vb{r}_{ij}\tran
  \braket*{\psi^+_n}{j}\!\!
  \braket*{i}{\psi^+_m}
  ,
  \\
  {j_a}_{nm}
  &=
  - {j_a}_{mn}
  =
  \I t
  \sum_{\nn{ij}} {\vc{r}_{ij}}_a
  \braket*{\psi^+_n}{j}\!\!
  \braket*{i}{\psi^+_m}.
\end{align}

\section{Spreading of correlations after an interaction quench}
\label{AppendixB}

\begin{figure}[t]
  \includegraphics[width=\linewidth]{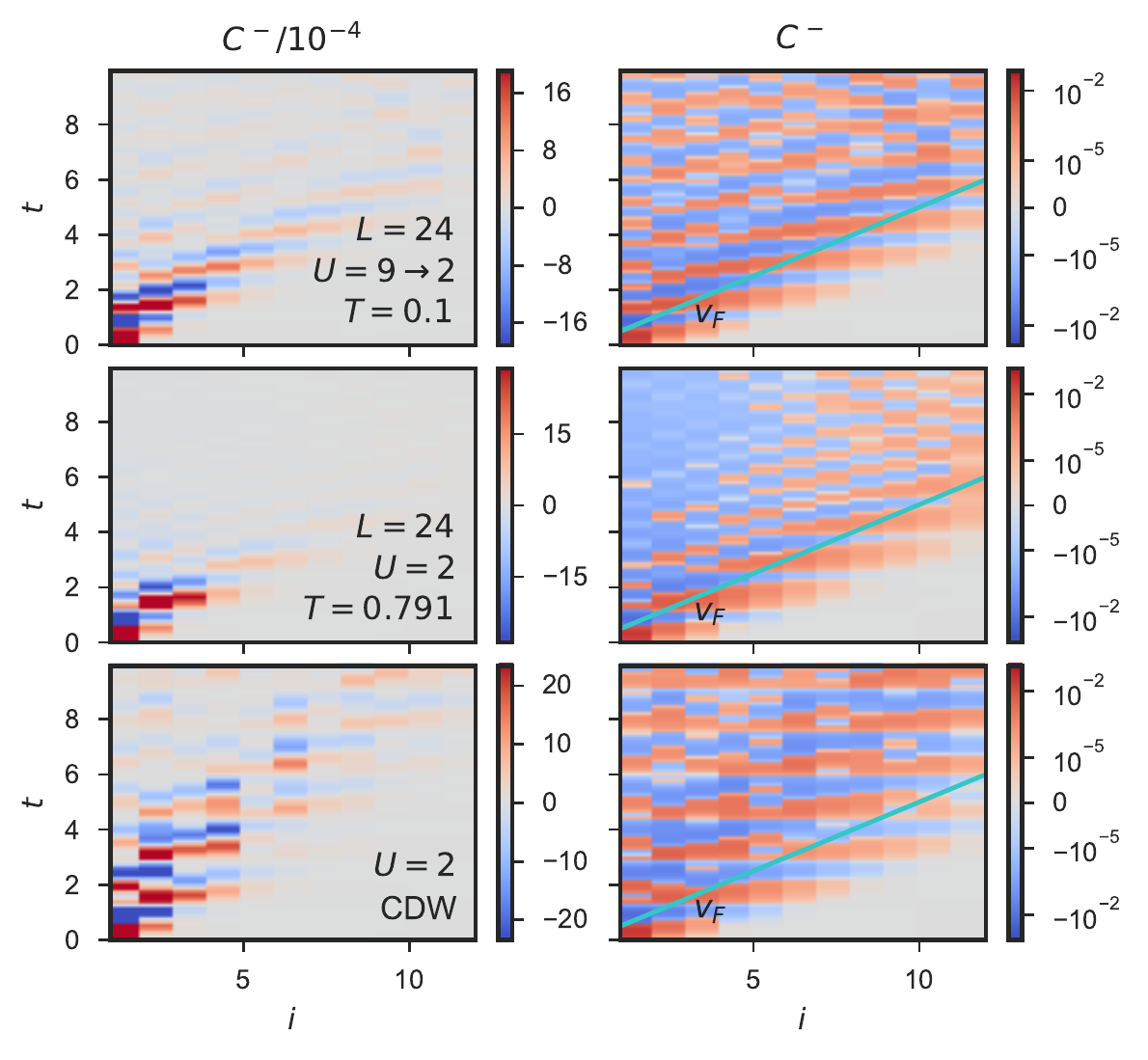}
  \caption{\label{fig:quench_and_effective_temperature}%
    Two-dimensional model: Spreading of two-point density-density correlations
    in a nonequilibrium system after an instantaneous interaction quench
    from $U=9$ to $U_q=2$ starting from an equilibrium system
    at temperature $T=0.1$ (top row).
    Spreading in the corresponding equilibrium system
    at the temperature $T_\mathrm{eff}$ defined in the text
    (center row).
    Spreading for a fixed $f$~particle configuration
    in a perfect charge-density-wave pattern.
    (bottom row).
  }
\end{figure}

To investigate the spreading of correlations in a nonequilibrium situation,
we consider an interaction quench in the two-dimensional model
from $U=9$ to $U_q=2$,
as described in \cref{eq:quench1,eq:quench2},
and an initial equilibrium temperature of $T=0.1$ (CDW phase).
The energy injected into the system
can be obtained from the instantaneous change in the local energy,
\ie\ the interaction and chemical potential contribution.
From the total energy of the system after the quench,
one may then calculate an effective temperature $T_\mathrm{eff}$,
which corresponds to the temperature of an equilibrium system
with $U=U_q$ and the same total energy.
For the above set-up, one finds $T_\mathrm{eff}=0.791$.\footnote{%
The Falicov-Kimball model is not expected to thermalize after the quench,
but this effective temperature can serve as a reference
to discuss the properties of the nonequilibrium state.}

In \cref{fig:quench_and_effective_temperature}
we plot the spreading of charge correlations
after the interaction quench (top panels)
and compare the result to an equilibrium simulation
at $U_q$ and $T_\mathrm{eff}$ (middle panels).
The spreading velocity is somewhat lower in the quenched system
than in the thermal system at $T_\mathrm{eff}$,
and the correlations are less confined.
This indicates an effectively ``cold'' (trapped) state of the quenched system.
In fact, comparing to the $T=0.1$ data in \cref{fig:spreading_correlations},
it seems that the correlations spread in a way
analogous to the cold CDW system at $U=U_q$,
despite the much higher total energy after the quench.
This is possible because the $f$~particle distribution
of the initial ($U=9$) CDW state
cannot adjust after the quench,
so that the disorder potential corresponds to a CDW state.

As further support of the dominant role of the disorder potential,
we have calculated the spreading behavior at $U=U_q$ and $T_\mathrm{eff}$
for a fixed $f$~configuration in a perfect CDW pattern (bottom panels).
This system shows no localization, despite the elevated temperature.
We also notice that the spreading velocity in the CDW potential
is reduced compared to the quenched system and the equilibrium system at $U_q$.
This shows that CDW correlations in the disorder potential
reduce the spreading velocity $v''$.

\vspace{40mm}
\mbox{}

\nocite{tange_gnu_2011}
\bibliography{literature}

\end{document}